# Development of UMLS Based Health Care Web Services for Android Platform

N. SOOMRO**, S. SOOMRO++, Z. ALANSARI, S. ABBASI, M. R. BELGAUM, A. B. KHAKWANI*

College of Computer Studies, AMA International University, Bahrain

**Abstract:** In this fast developing world of information, the amount of medical knowledge is rising at an exponential level. The UMLS (Unified Medical Language Systems), is rich knowledge base consisting files and software that provides many health and biomedical vocabularies and standards. A Web service is a web solution to facilitate machine-to-machine interaction over a network. Few UMLS web services are currently available for portable devices, but most of them lack in efficiency and performance. It is proposed to develop Android-based web services for healthcare systems underlying rich knowledge source of UMLS. The experimental evaluation was made to analyse the efficiency and performance effect with and without using the designed prototype. The understandability and interaction with the prototype were greater than those who used the alternate sources to obtain the answers to their questions. The overall performance indicates that the system is convenient and easy to use. The result of the evaluation clearly proved that designed system retrieves all the pertinent information better than syntactic searches.

**Keywords:** Unified medical language system, Web service, Health care system.

## 1. INTRODUCTION

Web service is the exchange of systems that contain objects, programs, documents, and a message that is used for application to application interaction through the internet. Two general approaches are for the development of web service. The first one is the top-down approach that is based on the web service interface and XML types. The implementation of the web service that create a WSDL file then use the Web Service to create the web service and skeleton to which can add the required code. The later approach is known as a bottom-up approach that is built on the basis of existing business logic. A WSDL file is generated to describe the resulting web service interface. It is often used for exposing existing function as a web service. However, if complex objects are used, the resulting WSDL might be difficult to understand and less interoperable (Keen, *et, at*. 2011).

The web services would also be useful to people who have certain symptoms but are unaware of the disease that they may be suffering. The proposed prototype is beneficial for the people specially suffering from Heart diseases, Diabetes, and bone disease, where every new research and discovery brings a new hope for life. This study describes the high potential of the interrelationship of information, people, and technology for improving the health care. The research discusses several information challenges associated with diabetes. This study suggested addressing a set of problems that will improve the lives of not only the patients but also their families, it will also make the provision of diseases care more effective and cost efficient. The major component of the prototype system is web services brain that is populated with the required terms.

## 2. BACKGROUND

Healthcare is the medical information system that includes health promotion and protection, disease prevention, health assessment, and illness surveillance of any targeted group. Healthcare systems are very essential for solving the growing medical problems by reducing the cost to find the exact solution, improving efficiency and quality of the solution. The Unified Medical Language System (UMLS) (Chen *et, al*. 2002), facilitates information integration and retrieval from multiple biomedical information sources using various vocabularies. It is preferred over MeshTree, due to huge number of concepts and terms with number of terms at each level of hierarchy in Mesh Tree (Soualmia, *et, al*. 2013). The UMLS has set of files and software that brings together many health and biomedical vocabularies and standards to enable interoperability between computer systems. Major components or knowledge sources of UMLS are shown in **(Fig-1).**

The Metathesaurus (Ruiz, *et, al*. 2012) is a compiled list of terms from over 40 autonomously controlled medical vocabularies known as source vocabularies. The Metathesaurus links synonymous terms from among these source vocabularies. The SPECIALIST Lexicon contains lexical variations of medical conditions in addition to non-medical words. Between them, the Metathesaurus and SPECIALIST Lexicon

++Corresponding author: Safeeullah Soomro, email: s.soomro@amaiu.edu.bh
* Department of General Studies, Jubail Industrial College KSA
**Indus University Pakistan

contain all of the terms that are documented in the UMLS.

UMLS is used to interconnect, drug names, billing codes, health information, and medical terms through unlike computer systems. In addition to that it includes terminology research, retrieval of search engine, statistical reporting of public health, and data mining (Mirhaji, *et, al.* 2005).

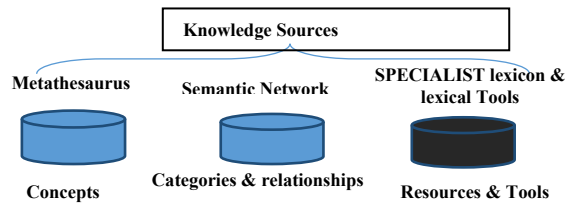

Fig-1 Knowledge sources of UMLS

## 3. MATERIALS AND METHODS
### 3.1 Design and Principles of System

This research is based on the objective to provide medical users with the terms base web service where users can interact with the system quickly. Various strategies and techniques for developing web service were identified by the literature review. The prototype was designed for the support of the thesis with the help of principles and techniques of web service using Android based application.

An information retrieval web service can replace traditional information retrieval systems such as Google as they provide direct answers to users' questions. The proposed solution that is an information retrieval web service reduces the time to browse and search through the documents to find the required information about the disease.

The measure hurdles for successful search in the medical field is the use of different vocabularies by various databases. Whereas, a query is frequently answered by the use of more than one database and mostly the users are unfamiliar with those vocabularies that lead to failed search due to the use of inappropriate terms. Because of this problem the searcher has to make numerous attempts to get best results to their queries.

This problem could be better solved by the use of UMLS meta-thesaurus as it is composed of concepts from a variety of vocabularies and classifications employed in the field of medicine. Along with keeping hierarchical information about concepts, UMLS have added new relationships between concepts from different vocabularies.

### 3.1.1 Metathesaurus

Metathesaurus is the nucleus of database, which is a collection of concepts and terms from various controlled vocabularies to link substitute names and views of the same concept from different source vocabularies and recognize meaningful relationships between various concepts. The vocabularies are either in Rich Release Format (RRF) or Original Release Format (ORF). It used to link the other UMLS Knowledge Sources.

### 3.1.2 Semantic Network

Semantic Network comprises semantic types and semantic relationships: Semantic types are Disease or Syndrome or Clinical Drug. The relationships of semantic exist between semantic types. For instance, Clinical Drug Treats Disease or Syndrome. Semantic Network is being used in several applications to help interpret meaning applied to the Metathesaurus concept.

### 3.1.3 Specialist Lexicon and Lexical Tools

Specialist Lexicon and Lexical tools are used to allow the users to develop Natural Language Processing programs. Lexicon includes many words from the biomedical domain. These words or terms are selected for lexical coding from many sources including MEDLINE and the general English vocabulary. It also describes the morphologic, orthographic and syntactic properties of a text. The lexical tools contain Java programs to process natural language words and terms as well as normalizer, a lexical variant generator and a word index generator (Mccray, *et, al.* 2001).

### 3.2 Web Service

The web service is a unit of managed code that can be remotely invoked by applying HTTP requests. Web service gives access to expose the functionality of available system over the computer network like the internet. When the system is presented on to the network, the other application may use the feature of the program. Web service offers implementation and technology independent platforms. The Web Service can be used to implement in existing low-cost internet or on other transport mechanisms like FTP etc. (Stutte, *et, al.* 2012).

Web Services Description Language (WSDL) is XML-based language used to describe web services and how to access them. Web service can read and interpret WSDL file to learn regarding the service location and operations (Bedrick, *et, al.* 2011).

Simple Object Access Protocol (SOAP) is used to send messages to a service and client. SOAP messages can be sent between applications despite of their platform that is interoperability (Grosjean, *et, al.* 2013). A lightweight platform and language neutral communication protocol allows programs to converse using Internet.

Universal Description, Discovery and Integration (UDDI) offer a mechanism for clients to discover other web services dynamically. UDDI is a service of directory where businesses can register and web services for search. UDDI is a public registry that one can publish and inquire about web services.

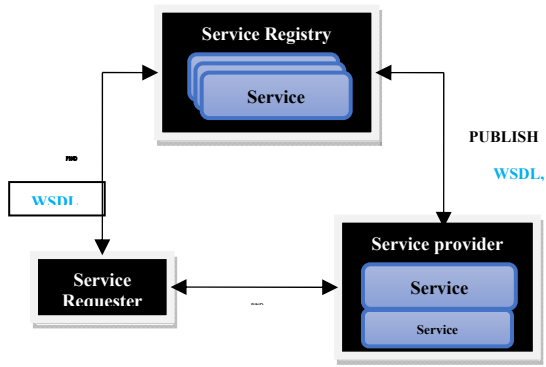

**Fig-2 Web Service Architecture**

### 3.3 System architecture

The System architecture is designed to map android base web services with underlying UMLS as knowledge-based systems. The system works as follows:

Initially an Android User has to place any medical based search query via Internet Connection by consuming a designed web service. The Web service will request the search term in UMLS knowledge-based System. If the required condition is found, the response will be given to the Android user by accessing particular web service.

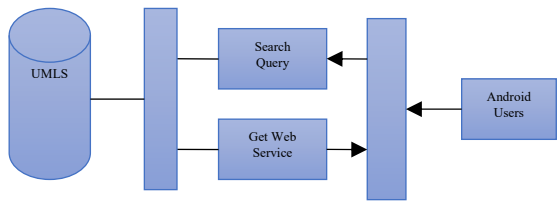

**Fig-3 System Architecture**

### 3.4 Designed prototype

Android Based Prototype is developed by implementing the UMLS as knowledge brain in order to overcome the issues addressed. The GUI designed requires basic details about Patients and its medical history, the details about any medical term could also be retrieved exactly (Wang, *et, al.* 2012). Screenshots of the designed prototype and its web service descriptions are given below:

The search for the Patients Details shown in **(Fig-4)** which is designed with simple and clear layout with understandable contents.

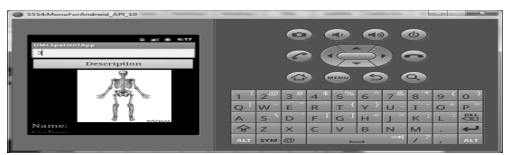

**Fig-4 Querying about the Patient using android application**

The HTTP POST method searches the record of specific patient with the details of the patient as shown in **(Fig-5)**.

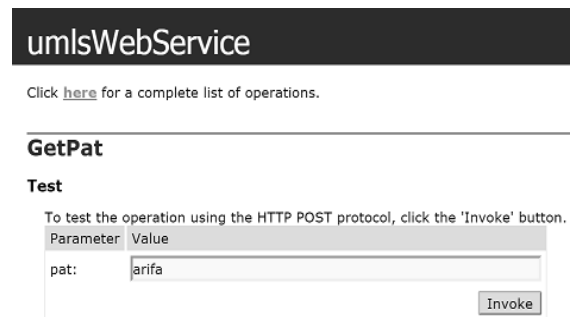

**Fig-5 Underlying Web service response**

The SOAP request and response after submitting the query related to the specific patients is **(Fig-6)**. The SOAP request message includes the SOAP envelop, which consists body holding the string for searching the patient ID. The SOAP response includes, SOAP Code, 200 Ok in case of successfully retrieved the code and the response message.

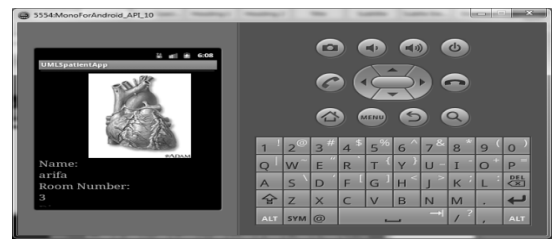

**Fig-6 Details about Patient**

The detail description about the medical terms is shown in the **( Fig-7)**.

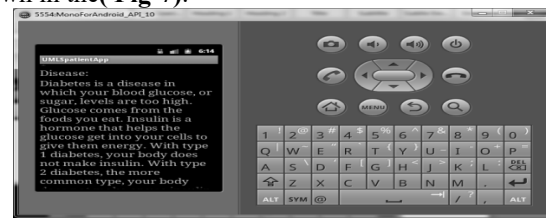

**Fig-7 Detailed description about the medical terms**

The detail descriptions about the medical terms are retrieved in the form of XML parser when the web service is invoked at the back end. The portion of XML Parser of designed web service is shown in **(Fig-8)**.

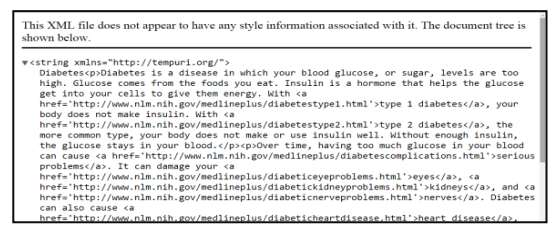

**Fig-8 XML Parser of retrieved web service**

## 4. DISCUSSION

A survey was conducted on a group of users with the use of prototype designed against use of alternate sources. **(Fig-9)** shows user Satisfaction results, which indicates the proposed web service prototype is overall close to satisfactory 34% and alternate source overall close to satisfactory 30%.

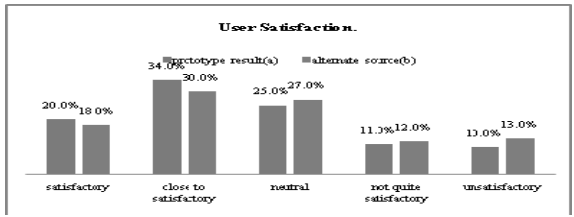

**Fig-9 Results of the survey regarding the User satisfaction**

**Fig-10** shows the results of interaction with the system. It indicates the proposed web service prototype is 26% close to satisfactory & alternate source is not quite satisfactory 26%.

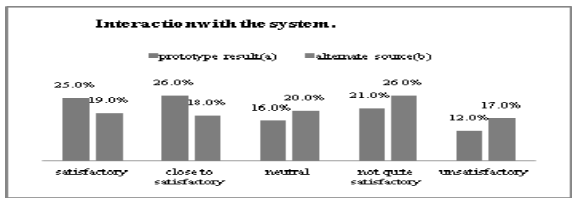

**Fig-10 Results of the survey regarding the interaction with the system**

**Fig-11** shows the Completeness of retrieved information. It shows that the proposed web service prototype is 35% satisfactory & alternate is 27% satisfactory.

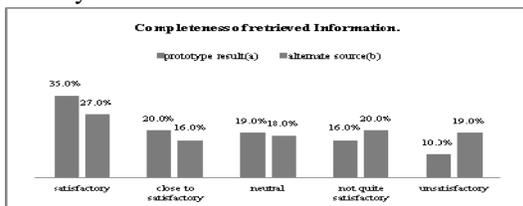

**Fig-11 Survey result of the Completeness of retrieved information**

**Fig-12** shows results of the System Understandability. It shows the proposed web service prototype is 25% neutral & alternate is 26% neutral.

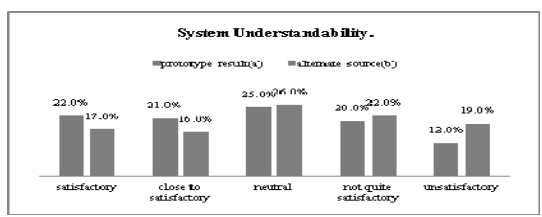

**Fig-12 Survey results of the System Understandability**

**Fig-13** shows the overall performance that the proposed web service prototype is 25% satisfactory & alternate 24% unsatisfactory.

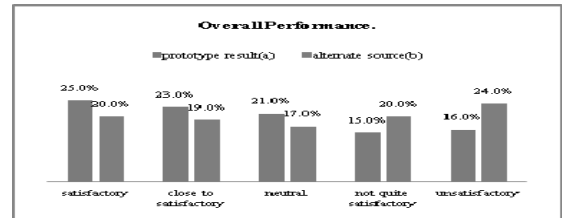

**Fig-13 Survey results of the overall performance**

## 5. CONCLUSION

The amount of knowledge in the medical domain is growing exponentially. With this growth, it is becoming very hard for the patients to keep track of all the new discoveries. The designed system addresses this issue and makes this knowledge development easier. The developed system performs Health Care Web Services by implementing UMLS as knowledge based system. It can be used by patients as well to discover resources related to their Personal Health Record.